\pdfoutput=1
\documentclass[manuscript]{emulateapj}

\usepackage{amsmath}
\usepackage{amssymb}
\usepackage[parfill]{parskip}   
\usepackage{ifpdf}
\usepackage{graphicx} 
\usepackage{longtable}
\usepackage{rotating}
\usepackage{epsf}
\usepackage{bm}
\usepackage{subfigure}
\usepackage{morefloats}
\usepackage{gensymb}
\usepackage{verbatim}
\usepackage{array}

\usepackage{ragged2e} 
\usepackage{color}
\usepackage{soul}

\newcommand{\HI}{{\sc Hi}}

\newcommand{\Oiii}{{\sc Oiii}}

\newcommand{\Nii}{{\sc Nii}}
\newcommand{\Sii}{{\sc Sii}}

\graphicspath{{figures/}}

\begin{document} 

\title{Choirs HI Galaxy Groups: ~\\The metallicity of dwarf galaxies}
\author{Sarah M. Sweet\altaffilmark{1}\altaffilmark{2}}
\author{Michael J. Drinkwater\altaffilmark{2}}
\author{Gerhardt Meurer\altaffilmark{3}\altaffilmark{4}}
\author{Kenji Bekki\altaffilmark{3}\altaffilmark{4}}
\author{Michael A. Dopita\altaffilmark{5}\altaffilmark{6}\altaffilmark{7}}
\author{Virginia Kilborn\altaffilmark{8}}
\author{David C. Nicholls\altaffilmark{5}}

\altaffiltext{1}{sarah@sarahsweet.com.au}
\altaffiltext{2}{School of Mathematics and Physics, University of Queensland, Qld, 4072, Australia}
\altaffiltext{3}{School of Physics, University of Western Australia, 35 Stirling Highway, Crawley, WA, 6009, Australia}
\altaffiltext{4}{International Centre for Radio Astronomy Research, ICRAR M468, 35 Stirling Highway, Crawley, WA, 6009, Australia}

\altaffiltext{5}{Research School of Astronomy and Astrophysics, Australian National University, Cotter Rd., Weston ACT 2611, Australia }
\altaffiltext{6}{Astronomy Department, King Abdulaziz University, P.O. Box 80203, Jeddah, Saudi Arabia}
\altaffiltext{7}{Institute for Astronomy, University of Hawaii, 2680 Woodlawn Drive, Honolulu, HI 96822, USA}
\altaffiltext{8}{Swinburne University of Technology, Mail number H30, PO Box 218, Hawthorn, Victoria 3122, Australia}

\date{\today}

\begin{abstract} 
We present a recalibration of the luminosity-metallicity relation for gas-rich, star-forming dwarfs to magnitudes as faint as M$_R\sim$ -13.
We use the \citet{Dopita2013} metallicity calibrations to calibrate the relation for all of the data in this analysis.
In metallicity-luminosity space we find two sub-populations within a sample of high-confidence SDSS DR8 star-forming galaxies; 52\% are metal-rich giants and 48\% are metal-medium galaxies. Metal-rich dwarfs classified as tidal dwarf galaxy (TDG) candidates in the literature are typically of metallicity 12 + log(O/H) = 8.70 $\pm$ 0.05, while SDSS dwarfs fainter than M$_R$ = -16 have a mean metallicity of 12 + log(O/H) = 8.28 $\pm$ 0.10, regardless of their luminosity, indicating that there is an approximate floor to the metallicity of low luminosity galaxies. Our hydrodynamical simuations predict that TDGs should have metallicities elevated above the normal luminosity-metallicity relation. Metallicity can therefore be a useful diagnostic for identifying TDG candidate populations in the absence of tidal tails. 
At magnitudes brighter than M$_R\sim$ -16 our sample of 53 star-forming galaxies in 9 H{\sc i} gas-rich groups is consistent with the normal relation defined by the SDSS sample. At fainter magnitudes there is an increase in dispersion in metallicity of our sample, suggestive of a wide range of H{\sc i} content and environment. In our sample we identify three (16\% of dwarfs) strong TDG candidates (12 + log(O/H) $>$ 8.6), and four (21\%) very metal poor dwarfs (12 + log(O/H) $<$ 8.0), which are likely gas-rich dwarfs with recently ignited star formation.
\\
\end{abstract}

\section{Introduction} 
Over the past decades, it has been shown that galaxies display an increasing metallicity with luminosity, and more fundamentally, mass \citep[e.g.][]{Lequeux1979,Garnett1987,Skillman1989,Brodie1991,Zaritsky1994,Tremonti2004,Zahid2012}.
The general explanation for this relation is 
 {that two concurrent processes are at work}. The first is that in the lambda cold dark matter framework, most galaxies start at the low luminosity, low metallicity end of the trendline; forming out of pristine gas clumping within dark matter (DM) haloes. Over time, DM haloes and their respective galaxies merge together to form ever-larger haloes and galaxies. The second, concurrent process is the self-enrichment of galaxies due to supernovae, with more massive galaxies retaining greater percentages of the ejecta than low-mass galaxies in the face of galactic winds \citep{Gibson1997,Kauffmann2003}, and/or being more efficient at converting the enriched gas into stars \citep{Dalcanton2007}. These two processes cause galaxies to move diagonally upwards along the trendline simultaneously towards higher mass and higher metallicity. 

However, not all dwarf galaxies are formed out of metal-poor gas in their own DM halo. Tidal interactions between giant galaxies cause knots of star formation in tidal tails, which can self-gravitate without the need for a DM halo. The dwarf galaxies formed in this way are known as tidal dwarf galaxies (TDGs), and have high metallicity due to the pre-enriched matter from which they form \citep[e.g.][]{Mirabel1992,Duc2000,Weilbacher2003}. There are a number of important implications for the study of TDGs, for example: the DM fraction within TDGs can constrain theories of gravity; the fraction of dwarf galaxies that form tidally and survive to the present day significantly affects the dwarf galaxy mass function \citep[see e.g.][]{Bournaud2010}.

It is not yet known what fraction of dwarf galaxies are TDGs; between `several' \citep{Bournaud2010} and 50 per cent \citep{Hunsberger1996} of dwarf galaxies are predicted to form in a tidal manner. This is still an open question, primarily because  two TDG criteria (low DM, high metallicity) are difficult to reliably quantify. Firstly, the presence of tidal streams around currently known TDGs prevents them from reaching the virial equilibrium required for measuring total mass and confirming the presence or absence of dark matter. Secondly, the various metallicity calibrations defined in the literature yield inconsistent metallicity measurements \citep[e.g.][]{Kewley2008}.

To test what fraction of dwarf galaxies form in a tidal manner, we have identified a sample of H{\sc i} gas-rich groups of galaxies where there is no obvious optical interaction, but the dwarf galaxies have higher rates of star formation than expected for the group environment. This sample is ideal for locating and measuring TDGs because the past interactions in the group provide the necessary conditions for the TDGs to form, but the lack of current optical interaction means that the TDGs would be old enough for their progenitor tidal tails to have dissipated since forming them, and the TDGs would be in virial equilibrium. This allows a sound measurement of their dynamical masses and tests of theories of gravity. 

Our aim is to determine the importance of tidal processes in forming dwarf galaxies in groups. In this paper we investigate the trend of metallicity with respect to luminosity of these objects in order to identify a population of candidate TDGs. 
{Here we define `metallicity' as the gas-phase oxygen abundance relative to hydrogen, 12+log(O/H). }The following section covers the sample selection, observations and data processing. In Section 3 we present the luminosity-metallicity relation, and discuss the implications in Section 4. Our conclusions are in Section 5.


\section{Sample Selection, Observations, Data Processing and Measurement}

Our sample consists of galaxies in small gas-rich groups named Choir groups \citep[][hereafter Paper 1]{Sweet2013}. The groups were selected from the HIPASS \citep[HI Parkes All-Sky Survey,][]{Barnes2001}, being the HI detections that were revealed by the Survey of Ionization of Neutral Gas Galaxies \citep[SINGG,][]{Meurer2006} to contain four or more emission line galaxies. In Paper 1, we presented a catalogue of the Choir group members and a discussion of their properties in the context of SINGG. Briefly, the Choir groups are on average more compact than groups in the \citet{Garcia1993} catalogue, but less so than Hickson compact groups \citep[HCGs;][]{Hickson1989}. Eight of them contain two large spirals and a number of dwarf galaxies, and as such are morphological analogues of the Local Group, albeit in a more compact state.

We observed 53 Choir member galaxies in nine groups with the integral field Wide Field Spectrograph \citep[WiFeS,][]{Dopita2007} on the Australian National University's 2.3m telescope. 
{This integral field unit has a 25'' x 38'' field of view with 1'' square spaxels.} The red R7000 and blue B3000 gratings (resolutions R = 7000 and 3000 respectively) were selected to achieve maximum 
 {velocity resolution of 45 km s$^{-1}$} in the red arm to facilitate redshift measurements with the H$\alpha$ line, and maximum sensitivity in the blue to facilitate measurement of abundance-sensitive spectral lines.  
 {The resulting wavelength ranges were 329-558 nm in the blue arm and 529-912 nm in the red.} The RT560 dichroic was used to ensure that the overlapping wavelength region did not contain strong features at the expected redshifts of our sample. Table~\ref{observations} lists the observing log. We found that the nod-and-shuffle observing mode provided the best sky subtraction, as it interleaves sky and object exposures to best account for time-varying sky brightness.
For each run we obtained the usual set of bias frames, and for each night a set of wavelength arc, flat and `wire' calibration frames. Spectrophotometric standard stars were observed nightly for each galaxy group. The data were processed using the IRAF-based pipeline described by \citet{Dopita2010}.

The best-known advantage of integral field unit (IFU) spectroscopy is the acquisition of spatially-resolved spectra. However, for this study we integrate over a number of spaxels (spatial pixels) per galaxy, so instead the advantages are increased signal to noise and an improved sampling over the entire galaxy. We are conducting a full spatially-resolved kinematic and metallicity analysis of these targets and will present the results in a future paper (Sweet et al., in prep).

For the dwarf galaxies, which fit within one pointing of the 25'' $\times$ 38'' field of view, we integrated over a grid of spaxels containing those with 
{$\gtrsim 3 \sigma$ signal in H$\alpha$}. This corresponds to 25-30 spaxels for a typical dwarf (in angular size) in our sample. For the giant galaxies, which do not fit within a single pointing, we integrated over the bright H{\sc ii} region nearest the centre of the galaxy. 
{We tested the effect of different aperture sizes on measured metallicity and found that expanding the aperture to include diffuse regions of the galaxies gave consistent metallicity results with those measured only for the bright central H{\sc ii} region. This corresponds well with previous findings that gas-rich, star-forming dwarfs are well-mixed }\citep{Kobulnicky1997,Lee2004}.
We measured redshifts for each integrated spectrum and confirmed that these are not background galaxies.

We measured emission line fluxes using {\sc uhspecfit} \citep{Rich2010}. This IDL-based program fits a \citet{Bruzual2003} stellar population to account for absorption,
before fitting Gaussian components for each emission line. 
{For most of the galaxies the integrated emission lines are narrow enough that a single-component Gaussian provides a good fit (see Figure 1); any residuals between the Gaussian fit and the observed spectrum are within the noise of the spectrum. For the giant galaxies that have broad components we have fit multiple Gaussian components, and again the residuals are within the spectrum noise.}
Reddening corrections were calculated based on the H$\alpha$/H$\beta$ ratio, assuming that the wavelength-dependent attenuation is due to an isothermal screen of dust, following \citet[][see their Appendix]{Vogt2013}.
Errors in flux measurements were estimated with a Monte Carlo simulation: simulated Gaussian distributions were added to random locations in the observed continuum, and the standard deviation of the measured fluxes was calculated.

\begin{figure}
\centerline{
\includegraphics[width=1\linewidth]{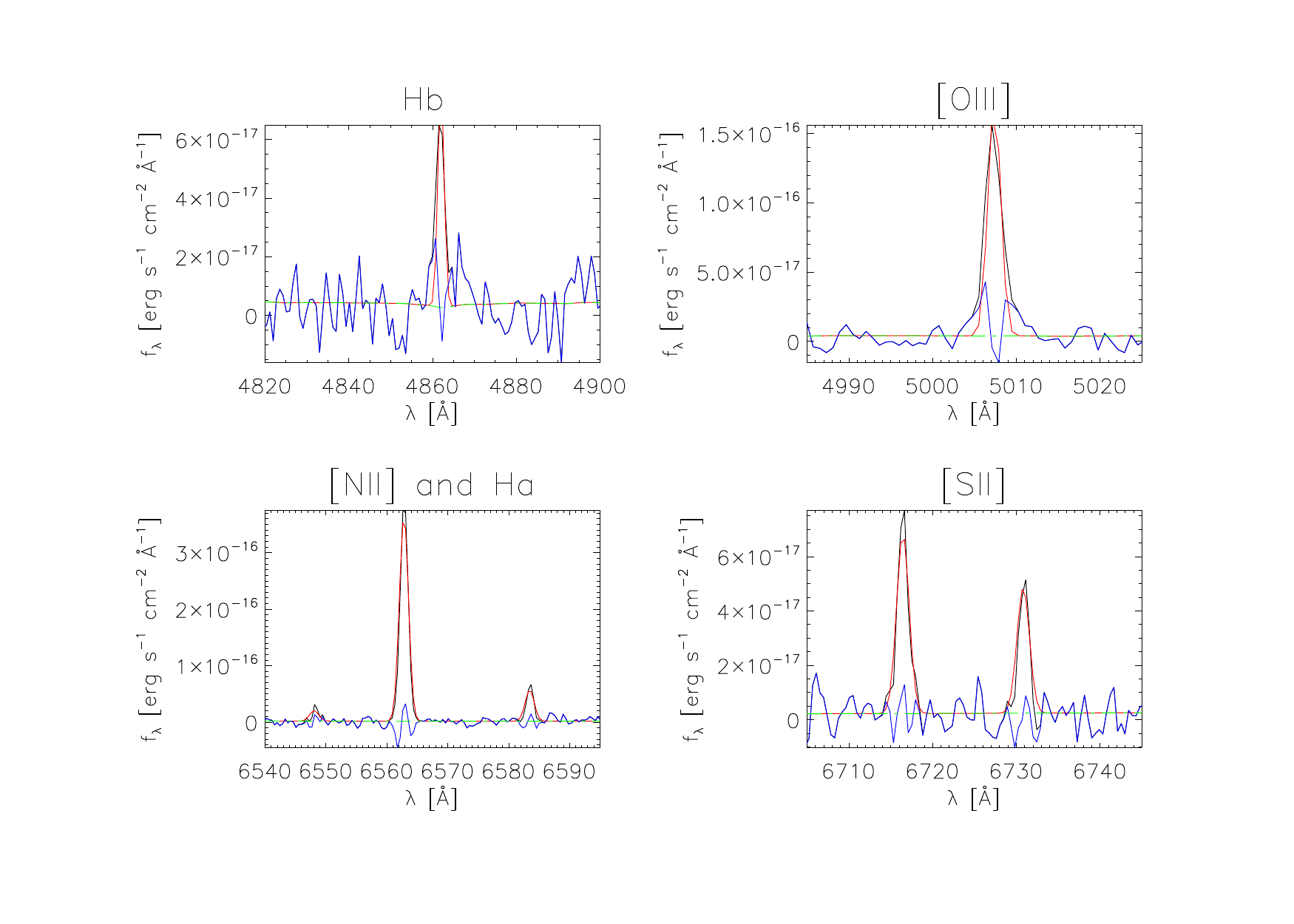}
}
\caption{Example of emission line fitting process. We show cutouts of one WiFeS spectrum (black) in the regions of interest, with best fit from UHSPECFIT (red), fitted continuum (green dashed), and residuals (blue). This spectrum is from HIPASS J1408-21:S5, the faintest dwarf in our sample.
\label{spec}}
\end{figure}

Example spectra are shown in Figure~\ref{spec}.
We present measured fluxes in Table~\ref{fluxes}.

\section{Results}

We constructed the luminosity-metallicity relation for our Choir member galaxies and comparison samples, using the same metallicity calibration (and where possible, reddening correction) for all of the measurements. Although the metallicity is more fundamentally related to 
{stellar mass than to luminosity} \citep[e.g.][]{Tremonti2004}, we restricted this analysis to luminosity because of the expected large scatter in the (unknown) mass-to-light ratios of our objects, which renders it difficult to make sensible, consistent mass estimates for these galaxies. In this section, we discuss our adopted metallicity calibration and discuss our comparison samples.

\subsection{Metallicity calibration}

Calibrations of gas-phase metallicity typically fall into three main categories:
\begin{enumerate}
\item{classical electron temperature and ionization correction factor technique \citep[e.g.][]{Peimbert1969,Stasinska1978,Stasinska2005
},}
\item{recombination line method \citep[e.g.][]{Esteban1998,
LopezSanchez2007
}, and}
\item{strong emission line (SEL) method \citep[e.g.][]{Pagel1979,McGaugh1991,Kewley2002,Dopita2013}.}
\end{enumerate}

Measuring electron temperature 
{ allows a} `direct' measurement of metallicity 
{from strongly temperature-dependent emission lines. As such it is} seen as the gold standard, but is difficult in practice because the required auroral lines (e.g. [O{\sc iii}] $\lambda$ 4363 \AA) are weak. Further, the auroral line strengths are anticorrelated to metallicity, so are only measurable for low-metallicity galaxies. The recombination line method 
{is also difficult, because the recombination lines are intrinsically weak.} These methods are therefore reserved for bright and/or nearby galaxies. The galaxies in our sample are mostly high-metallicity, faint and not very nearby, so most do not display the required lines for either the electron temperature or recombination line methods. 
{Although the SEL method has limitations, which are discussed in the following paragraph, it is better suited to our sample than the other two methods.} We therefore adopt the SEL method for this work.

Unfortunately, the three categories of methods give different results, so it is difficult to compare metallicities that have been calibrated with different methods. There is even wide variation within the various SEL methods, as seen in Figure 4 of \citet[][]{Kewley2008}. In part this is because the models are often degenerate: 
{many, such as the oft-used $R_{23}$ calibration} \citep{Pagel1979} 
{have} high- and low-abundance branch
{es, determined} by differences in ionization parameter $q$
{. This leads to} an undefined region below the degeneracy in metallicity \citep[][]{LopezSanchez2012}. The other major cause of discrepancy between the models is their failure to account for known physics. For instance, the models assume a Maxwellian photon energy distribution. \citet{Nicholls2012,Nicholls2013} 
{suggested that a} high-energy excess of ionizing photons 
{could be} characterised as a `$\kappa$-distribution' (generalized Lorentzian distribution).
\citet{Dopita2013} have since developed a SEL model which accounts for this distribution, 
{and encouragingly gives} much more consistent results with recombination line and electron temperature methods.

\begin{figure}
\centerline{
\includegraphics[width=1\linewidth]{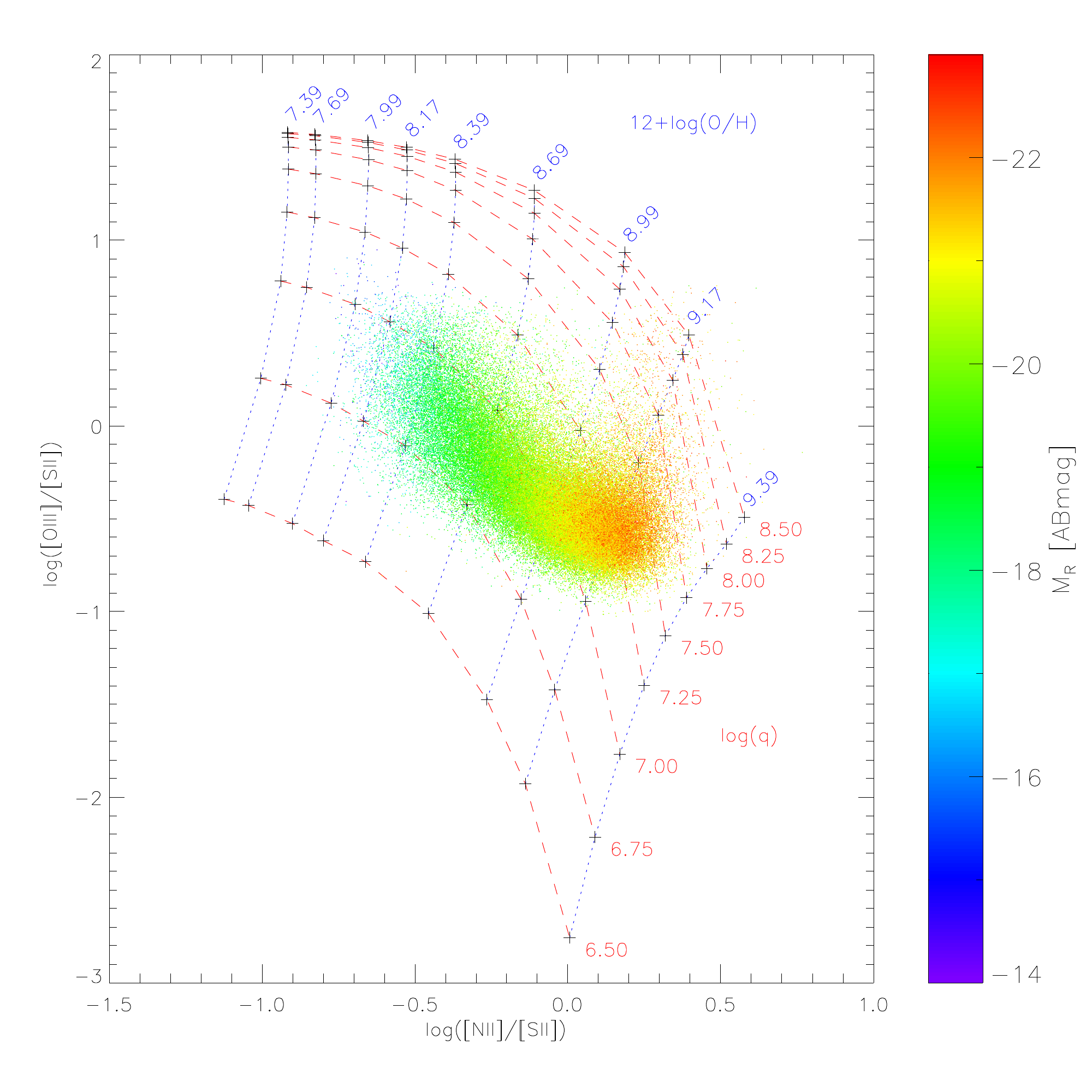}
}
\caption{The \citet{Dopita2013} metallicity calibration grid for [O{\sc iii}]/[S{\sc ii}] vs. [N{\sc ii}]/[S{\sc ii}], which 
{illustrates the metallicity calibration} for our sample. The red
{, dashed model curve labels depict} log(ionization parameter), while the blue
{, dotted model curve labels denote} the metallicity. Here we 
{show the} galaxies in our SDSS \citep{Aihara2011} sample, colour-coded by magnitude.
\label{diagnostic}}
\end{figure}

\begin{figure}
\centerline{
\includegraphics[width=1\linewidth]{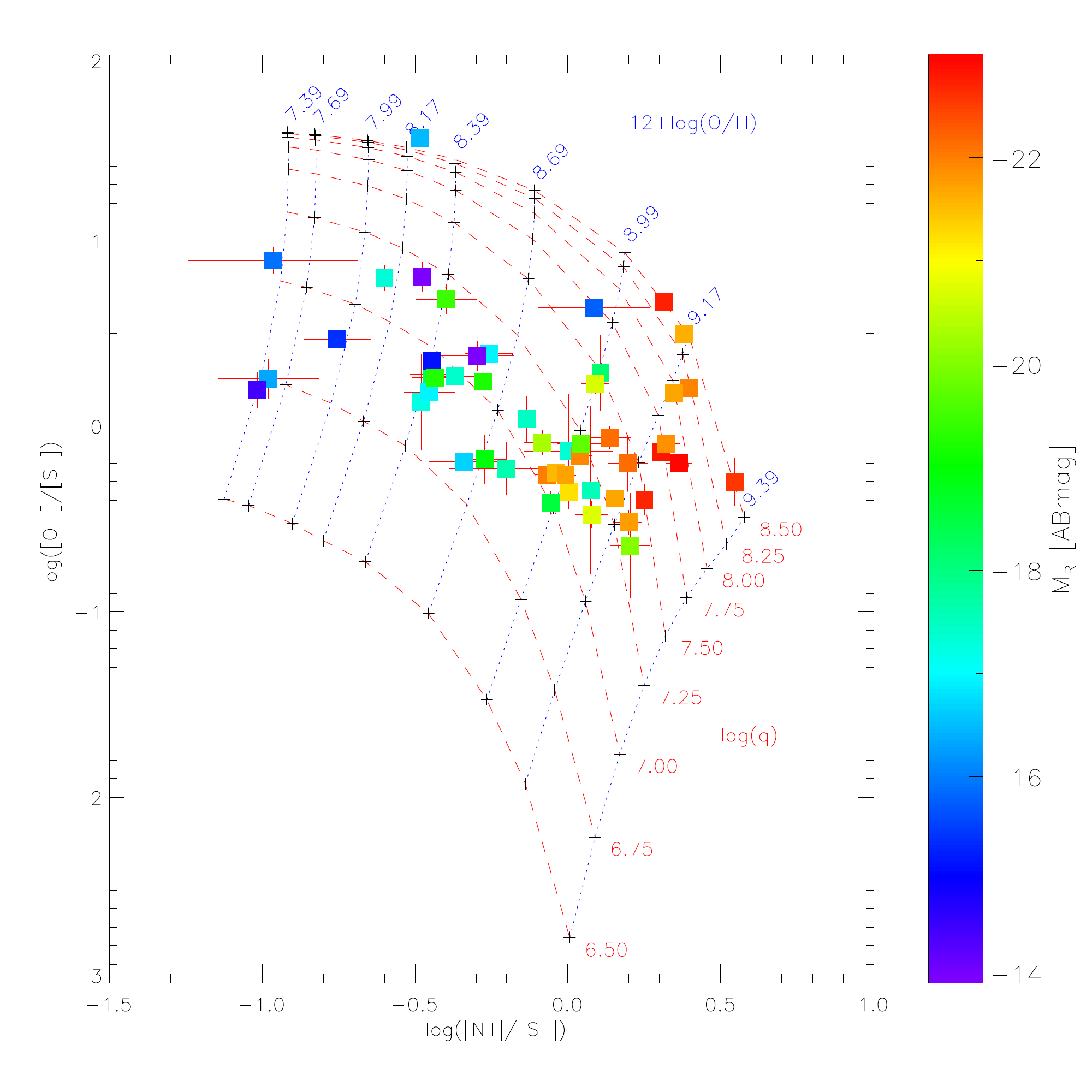}
}
\caption{As above, for our sample of galaxies in gas-rich Choir groups.
\label{diagnostic_choir}}
\end{figure}

{Furthermore}, when analysing metallicities by the SEL method, it is important to choose (i) a single metallicity calibration (so that the sample is self-consistent), that (ii) is as free of 
{degeneracy as possible}. For these reasons, we adopt the log [\Oiii]/[\Sii] vs. log [\Nii]/[\Sii] diagnostic given in \citet[][their Fig. 21 and our Figures~\ref{diagnostic} and~\ref{diagnostic_choir}]{Dopita2013}. The diagnostic is useful because it provides a clear separation of the ionization parameter $q$ and metallicity 12 + log(O/H), and is not highly dependent on the value of $\kappa$. Following \citet{Dopita2013} we adopt $\kappa$ = 20. We use a bivariate polynomial interpolation to convert the 
{diagnostic} grid from line ratio to ionization parameter - metallicity space, using {\sc rmodel} \citep{Cardiel2003}\footnote{http://www.ucm.es/info/Astrof/software/rmodel/rmodel.html}. 
Errors in metallicity for the Choir member galaxies are estimated by {\sc rmodel} with a Monte Carlo simulation based on errors in the emission line ratios. The non-regular shape of the calibration model means that the errors are asymmetric. As expected, the errors are generally larger for fainter galaxies, and where the metallicity is low (because the low-metallicity region of the grid is the most sensitive to [N{\sc ii}]/[S{\sc ii}]). 

Figure~\ref{diagnostic} illustrates our non-degenerate 
{metallicity diagnostic with our} SDSS control sample, which is presented in the following section. On that control sample, the \citet{Dopita2013} calibration is higher than the hybrid calibration method used in \citet{Tremonti2004} by 0.1-0.2 dex over the relevant magnitude range (see Fig~\ref{fits}). This difference was previously noted in \citet{LopezSanchez2012}. 
Figure~\ref{diagnostic_choir} also illustrates our metallicity 
{diagnostic}, but for our sample of galaxies.


\subsection{Control samples}

\subsubsection{SDSS}
Due to availability of quality photometry and spectoscopy for 860,000 galaxies, the Sloan Digital Sky Survey Eighth Data Release  \citep[SDSS DR8,][]{Aihara2011} is an ideal catalogue from which to draw our bright galaxy comparison sample. Following \citet{Tremonti2004}, we restrict our SDSS sample to a selection of high-confidence detections. The selection limits are as follows: 0.005 $<z<$ 0.25; $>$ 5$\sigma$ detection in each of H$\beta$, [O{\sc iii}], H$\alpha$, [N{\sc ii}], and [S{\sc ii}]; log([O{\sc iii}]/H$\beta$) $<$ 0.61/(log([N{\sc ii}]/H$\alpha$)-0.05)+1.3 \citep[to exclude AGN, following][]{Kauffmann2003}
; classified as galaxy; $\sigma_z< 0.15$; $\sigma_{H\delta} <$ 2.5; $\sigma_{Dn(4000)} <$ 0.1. These parameters ensure that our SDSS sample is consistent with the Tremonti sample, and clean of most spurious detections. Further, we visually inspected the 300 faintest (M$_R >$ -16 magnitude) galaxies and excluded 30 H{\sc ii} regions which were incorrectly classified as galaxies. Most of these have high metallicity ($\sim$ 8.6 - 9.0) corresponding to the parent galaxy, but all of them have faint magnitudes corresponding to the local H{\sc ii} region, so cannot be included in the luminosity-metallicity relation. Our resulting SDSS sample contains 94,863 sources.

We then converted SDSS $r$-band absolute magnitude to SINGG $R$-band AB absolute magnitude using the $r_{SDSS}$ to $R_{Vega}$ transformation by Robert Lupton\footnote{http://www.sdss.org/dr5/algorithms/sdssUBVRITransform.html} and the $R_{Vega}$ to $R_{AB}$ Deep Lens Survey transformation\footnote{http://dls.physics.ucdavis.edu/calib/vegaab.html}. We consider that the $r$-band SDSS to AB magnitude correction is small compared with the scatter in the $r_{SDSS}$ to $R_{Vega}$ conversion\footnote{http://www.sdss.org/dr5/algorithms/fluxcal.html\#sdss2ab}, so adopt the final conversion M$_{R(AB)}$ = M$_{r(SDSS)}$ - 0.1837($g_{SDSS}$ - $r_{SDSS}$) + 0.0829.

Using the methods described above, we performed reddening corrections and metallicity calibrations for this sample. 

\begin{figure}
\centerline{
\includegraphics[width=1\linewidth]{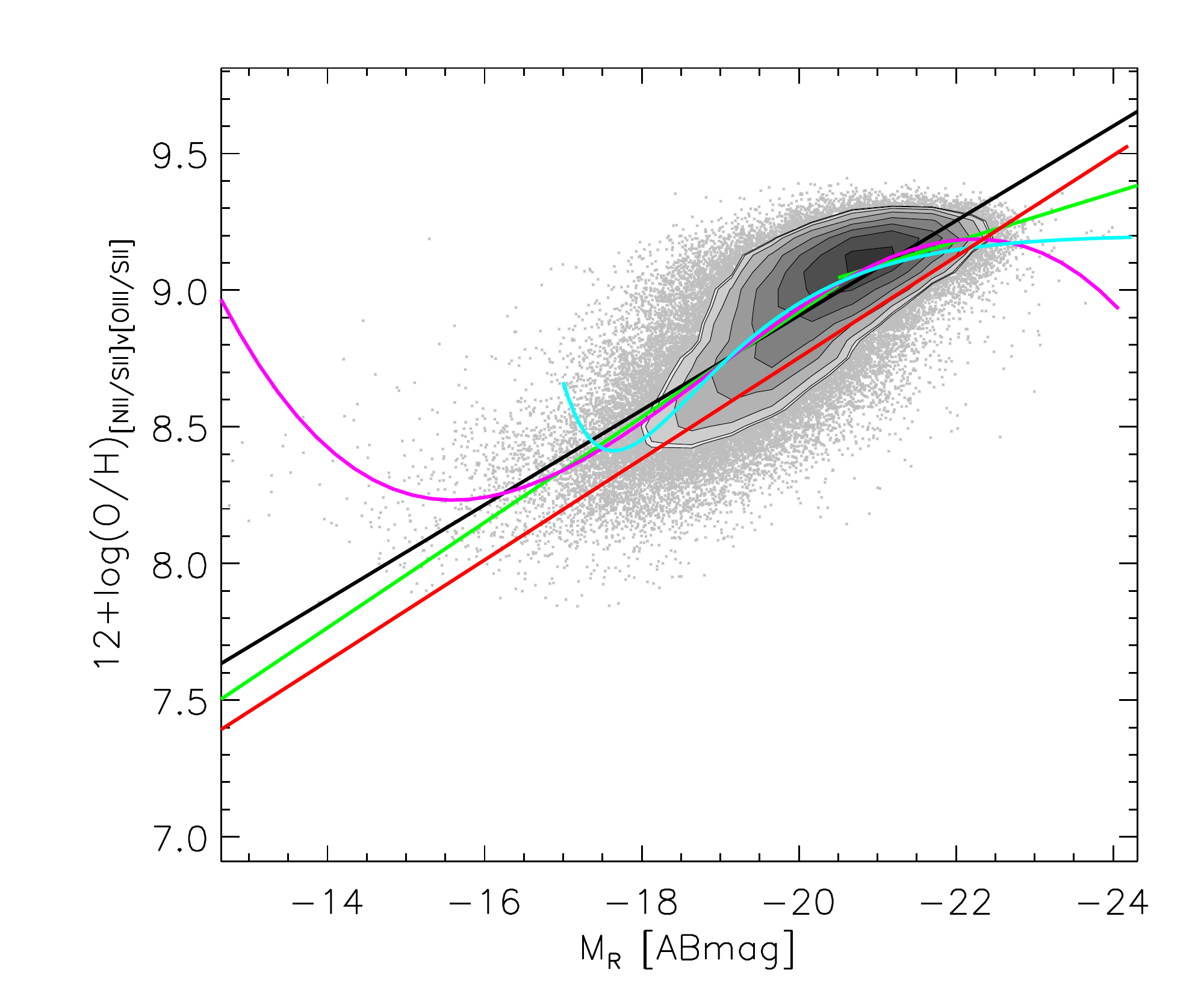}
}
\caption{Luminosity-metallicity relation for our SDSS control sample - grey points and grey-scale contours. We overlay various fits: linear=black, piecewise linear=green, cubic=magenta, asymptotic=cyan, Tremonti linear=red. The linear fit using the \citet{Dopita2013} calibration is 0.1-0.2 dex above the \citet{Tremonti2004} fit.
\label{fits}}
\end{figure}

We plot the luminosity-metallicity relation for our SDSS sample in Figure~\ref{fits}. 
 We attempt to model the luminosity-metallicity relation for SDSS using some common approaches in the literature: linear, piece-wise linear, polynomial, asymptotic fits \citep{Tremonti2004,Kewley2008,Mannucci2010,Sanchez2013}. It is qualitatively evident that none of these models fit the data very well, particularly at faint and/or very bright magnitudes. 

\begin{figure}
\centerline{
\includegraphics[width=1\linewidth]{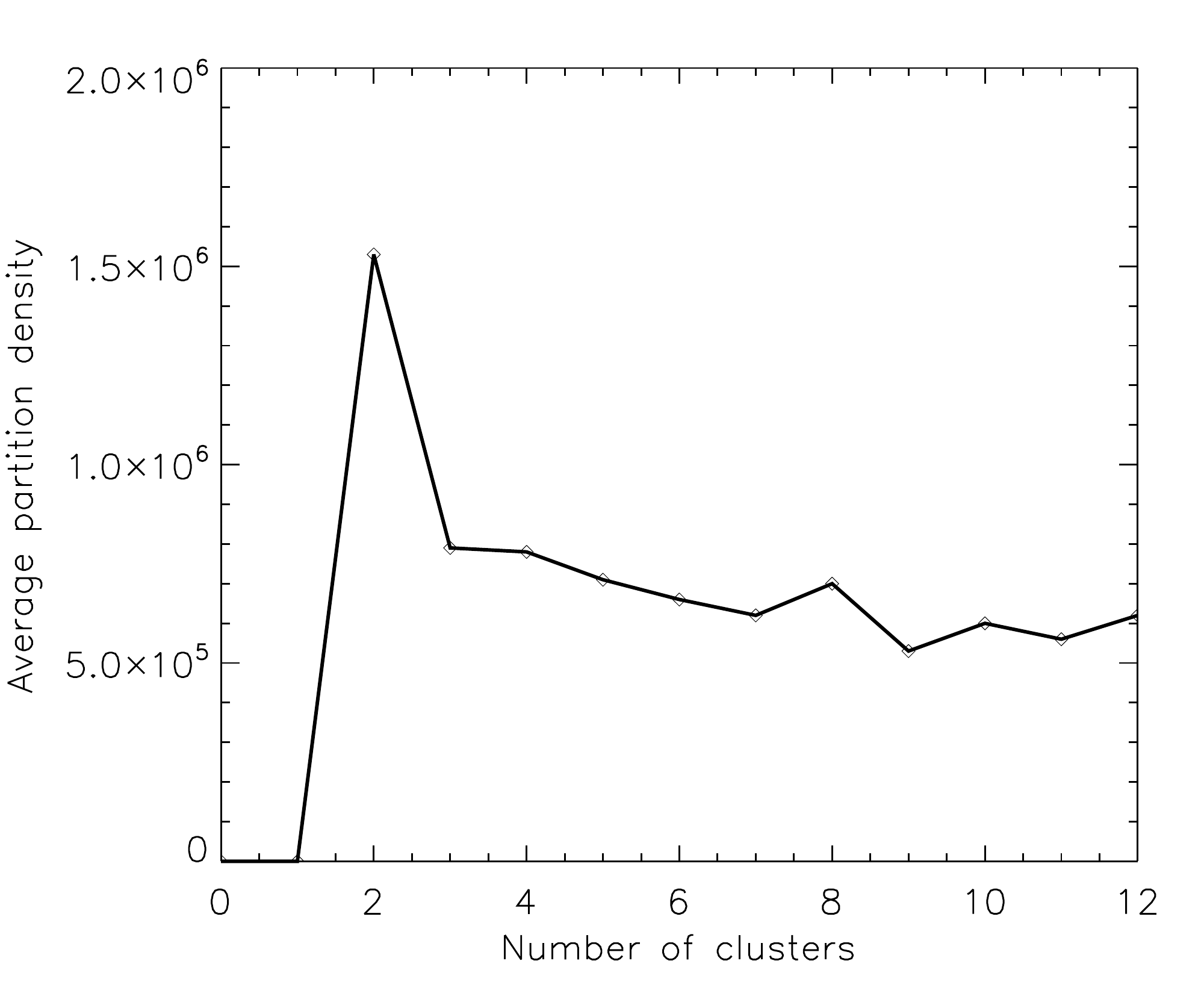}
}
\caption{Figure of merit showing goodness of fit as a function of number of sub-populations (`clusters') fitted for in our SDSS control sample. The goodness of fit is measured by average partition density, which is defined as the sum of memberships near centres divided by the volume of clusters, normalised to the number of clusters. The optimum number of clusters is clearly two.
\label{merit}}
\end{figure}

\begin{figure*}
\centerline{
\includegraphics[width=1\linewidth]{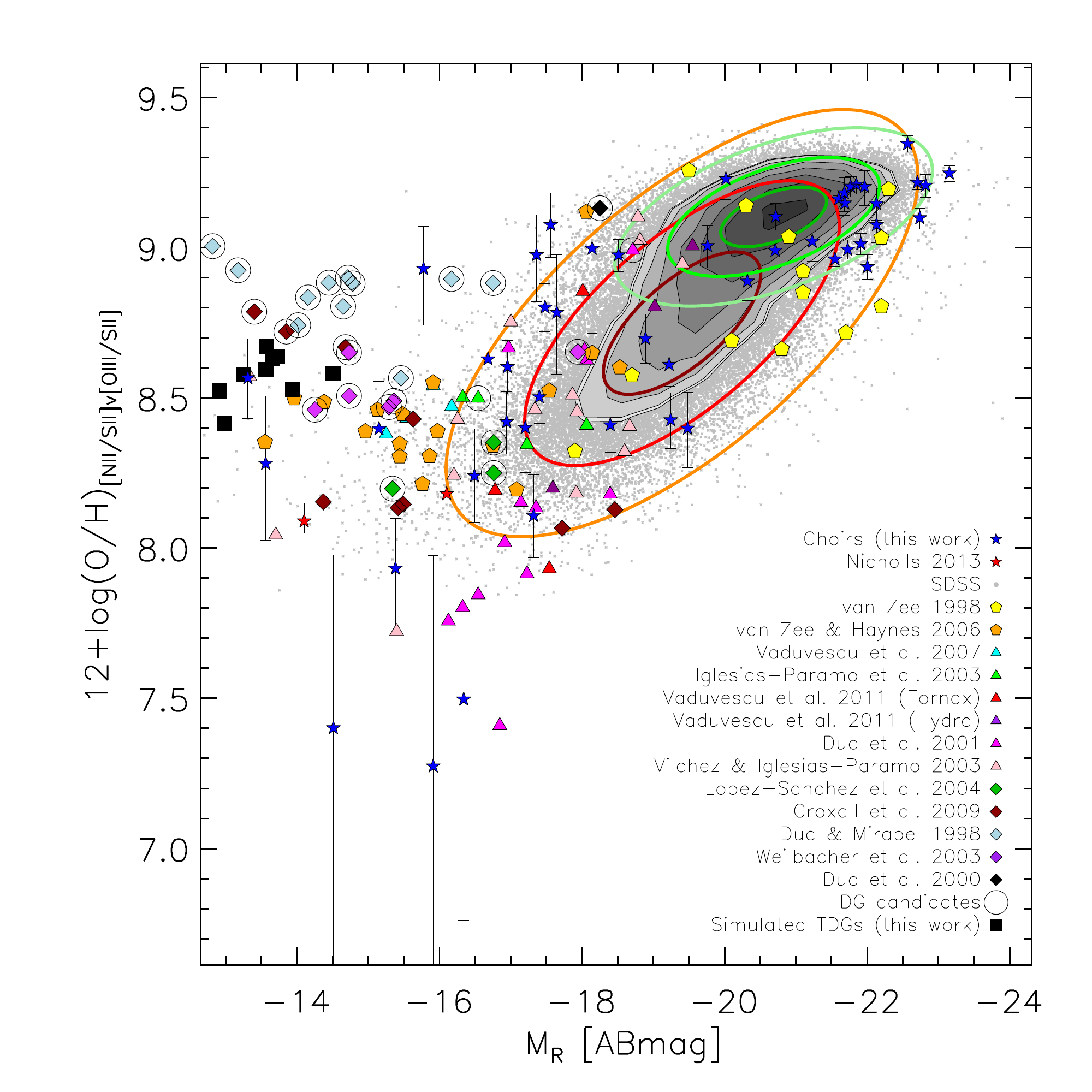}
}
\caption{\small Luminosity-metallicity relation for our SDSS control sample, with Gaussian mixture modelling overlaid; the two sub-populations are shown in red and green 1-, 2-, 3-$\sigma$ ellipses. Choir galaxies are shown in blue stars. Other dwarf galaxies in the literature are also shown; pentagons denote isolated galaxies, triangles denote gas-rich galaxies, and diamonds denote dwarf galaxies very near a host. Tidal dwarf galaxy candidates are circled; on average these are elevated above the normal relation defined by the SDSS sample. We also show our TDGs from our hydrodynamical simulations as black squares. Our Choir galaxies have a wide range in metallicity; three of these are significantly above the normal SDSS relation and are therefore strong TDG candidates.
\label{LZ}}
\end{figure*}

Moreover, a clear turnover, or knee, can be seen in the luminosity-metallicity relation. The poorness of the traditional fits together with the hint of multiple populations motivate us to perform Gaussian mixture modelling, which identifies sub-populations (`clusters') in multi-dimensional data with a maximum likelihood approach. In particular we use the unsupervised optimal fuzzy clustering algorithm described by \citet{Gath1989}
{, varying the number of clusters, $k$}. It is common to measure the goodness of fit by the density of each cluster (number of members near the centre of a cluster, divided by its total number of members), but as $k$ increases towards the sample size, the density of each cluster increases, so the fit becomes increasingly good. Instead, to avoid over-fitting, we calculate the `average partition density' as defined by \citet{Gath1989}, where cluster density is normalised to the number of clusters $k$. We plot this as our figure of merit in Fig.~\ref{merit}; a larger average partition density represents a better fit. Clearly, the optimum number of clusters is $k$=2. We plot the 1-, 2- and 3-$\sigma$ ellipses for these two sub-populations in Fig~\ref{LZ}. There is a metal-rich (12+log(O/H) = 9.1), `giant' (M$_R$ = -20.7) sub-population, containing 52\% of the sample, and a metal-medium (12+log(O/H) = 8.5), `medium+dwarf' (M$_R$ = -19.5) sub-population, with the remaining 48\%.  The overlap in luminosity-metallicity space suggests that there are other dimensions that may distinguish between the sub-populations, such as ionization parameter $q$. The fact that the knee is also seen in the calibration grid in Fig.~\ref{diagnostic} lends support to this idea. Here, low luminosity galaxies (blue-green; M$_R \lesssim$ -21) have increasing metallicity but fairly constant ionization parameter (log $q \sim$ 6.7-7), then brighter galaxies (yellow-red; M$_R\gtrsim$ -21) have increasing metallicity and increasing ionization parameter (up to log $q$ = 8). This combination of increasing metallicity and ionization parameter was also found in a spaxel analysis of a sample of luminous infrared galaxies by \citet{Dopita2013b}.

At magnitudes brighter than M$_R \sim -16$, SDSS is an ideal control sample due to the volume and quality of data (see Fig.~\ref{LZ}). At magnitudes fainter than M$_R \sim -16$, there are two possible concerns with the SDSS control sample, which we address here. (i) There is an apparent metallicity floor to the SDSS population, with no metallicities lower than approximately 12 + log(O/H) = 7.9. For SDSS dwarfs fainter than M$_R \sim -16$, metallicity is constant with luminosity, with a mean of 12 + log(O/H) = 8.28 $\pm$ 0.10. We note that this floor is also seen (albeit less obviously, due to different figure scale and limits) in \citet[][]{Tremonti2004} and Fig. 21 of \citet[][]{Dopita2013}. This floor is not an artifact of the metallicity calibration, because the calibration is well-defined down to 12 + log(O/H) = 7.39 and is not degenerate. Moreover, the floor is not absolute since we observe Choir (and other) galaxies below this metallicity, and these are calibrated using the same method. Finally, the metallicity floor cannot be explained by selection effects. 
{Although one may consider low surface brightness dwarfs, with few HII regions and low metallicity, to be selected against in our sample due to low signal-to-noise ratio, we see the same floor even without signal-to-noise cuts.} We point out that the floor could be an artefact the aperture effect in SDSS whereby faint, nearby objects are large in angular size compared with the fibres. The SDSS measurements are consequently of nuclear spectra for these galaxies, which are higher than the mean galaxy abundance due to galactic abundance gradients. 
We therefore consider for this analysis that it is probably a true lower limit for the nuclei of typical galaxies, and will investigate the floor further in a future work. The exceptions to this limit are discussed in Section \ref{discussion}.
(ii) There is an increased dispersion in metallicity at magnitudes fainter than M$_R \sim -16$. The SDSS sample is selected to only contain high-confidence detections, so the dispersion is a physical dispersion in the galaxies, rather than caused by measurement error. We conclude that the SDSS sample is therefore of sufficient quality to act as a control sample for our population of dwarfs fainter than M$_R \sim -16$.

\subsubsection{Additional dwarf galaxy control samples}

We now turn to a number of other samples for which [O{\sc iii}], [N{\sc ii}] and [S{\sc ii}] data are available. Where H$\alpha$ and H$\beta$ are also available and the observed fluxes are given we apply the same reddening correction as for the Choirs and SDSS samples. Where these are unavailable we adopt the dereddened fluxes given by the authors. Where necessary, we convert from $B$- or $K$-band magnitudes assuming typical colours $B-R\approx$ 1 and  $R-K\approx$ 2 \citep{Binney1998}.

We include two dwarfs from the small, isolated, gas-rich, iregular dwarf galaxy sample \citep[SIGRID,][]{Nicholls2011}, that have also been measured with the WiFeS integral field spectrograph. Full details are presented in \citet{Nicholls2013b}. We include additional H{\sc ii} regions and isolated dwarfs from \citet{VanZee1998} and \citet{VanZee2006}. For each galaxy in these three samples we sum over the emission line fluxes measured in all of the H{\sc ii} regions within that galaxy, before calculating line ratios and interpolating to metallicity as before. 
{Our tests show that this gives the same result as averaging the metallicities for each H{\sc ii} region, consistent to within $\pm$ 0.05 dex.} The integrated metallicities are then plotted against total galaxy luminosities.

The SIGRID dwarfs are plotted as red stars in Fig.~\ref{LZ}. Both of these (KK[98] 246 and HIPASS J1609-04) are consistent with the SDSS sample.

The \citet{VanZee1998} and \citet{VanZee2006} galaxies are shown as pentagons in Fig.~\ref{LZ}. The bright galaxies are consistent with the SDSS sample, but the faint end is elevated in metallicity above SDSS, at a constant metallicity with luminosity (12 + log(O/H) = 8.46 $\pm$ 0.04). We note that Fig. 21 of \citet{Dopita2013} also indicates a similar discrepancy between SDSS and the \citet{VanZee2006} sample. 
{At the low-metallicity end of the diagnostic, the metallicity is almost entirely determined by the [N{\sc ii}]/[S{\sc ii}] ratio. This means that metallicities are dependent on the assumed relationship between N/O and 12 + log(O/H). The N/O relationship has been recalibrated for the }\citet{Dopita2013}
{ model grids, so metallicities measured with these models will be offset from metallicities measured with earlier models. However, we have used the recalibrated model for all of the samples in this analysis, so this recalibration of N/O does not cause the elevation of the }\citet{VanZee1998} and \citet{VanZee2006}
{ samples over SDSS.}

We plot dwarf galaxies belonging to various local clusters as triangles in Fig.~\ref{LZ}: Virgo \citep{Vaduvescu2007,Vilchez2003}, Hercules \citep{Iglesias2003}, Fornax \citep{Vaduvescu2011}, and Hydra \citep{Vaduvescu2011,Duc2001}. In general, these objects tend towards lower metallicity with faint luminosity. \citet{Vaduvescu2007} 
{found that the H{\sc i} gas-richness of dwarfs has an effect on their metallicity. This is borne out by the general trend of these gas-rich cluster dwarfs (particularly the Hydra dwarfs in }\citet{Duc2001}
{) towards very low metallicities, compared with the isolated samples, which are not gas-rich.}

Finally, we also include a selection of dwarfs in groups, plotted as diamonds: NGC5291 \citep{Duc1998} and Arp245N \citep{Duc2000} (both in pairs), the compact group HCG31 \citep{LopezSanchez2004}, the larger $\sim$30-member group M81 \citep{Croxall2009}, and various other interacting systems \citep{Weilbacher2003}.

 \section{Discussion\label{discussion}}
 
 \subsection{Tidal dwarf galaxies}
Tidal dwarf galaxies are expected to have high metallicity corresponding to the pre-enriched material from which they form. They should also be formed within a tidal tail (which may or may not be sufficiently bright to observe), without a DM halo, and may be located near their parent giant galaxy (depending on the time since formation), \citep[e.g.][]{Duc2000,Weilbacher2003}. Here, we use the luminosity-metallicity relation to identify candidate TDG galaxies for later follow-up.

A number of the galaxies from the existing literature shown in Fig.~\ref{LZ} are claimed by their authors to be TDG candidates (circled points). As for the isolated dwarf sample, these TDG candidates do not display an increasing metallicity with luminosity, but they differ from the isolated dwarf sample by showing an enhanced average metallicity (12 + log(O/H) = 8.70 $\pm$ 0.05). While some of them are clearly elevated above the luminosity-metallicity relation of SDSS bright galaxies and \citet{VanZee2006} isolated dwarfs \citep[e.g. Arp245N, black diamond,][]{Duc2000}, many are consistent with the SDSS control sample \citep[e.g. HCG31 TDG candidates, dark green diamonds,][]{LopezSanchez2004}. The overlap between isolated dwarfs (non-TDGs) and previously-identified TDG candidates is partly due to the fact that some of those TDG candidates were identified because they have a higher metallicity than normal / isolated dwarfs using different metallicity calibrations. This overlap therefore simply confirms that 
{using different methods to measure metallicity will give different results.}

\subsection{Simulations}
We have conducted hydrodynamical simulations of TDG candidates, the full details of which will be given in Bekki et al. (in prep., 2013). Brief details follow. We model a Milky Way-type disk galaxy  with  a total dark halo mass of 10$^{12}$ M$_\odot$,
a stellar mass of 6 x 10$^{10}$ M$_\odot$, bulge mass of 10$^{10}$ M$_\odot$,
and gas mass of 3 x 10$^{10}$ M$_\odot$. The adopted initial 
{(stellar and gas-phase)} metallicity gradient is -0.08 dex kpc$^{-1}$, with nuclear 
{stellar} metallicity of [Fe/H]= 0.34 dex 
{and gas-phase metallicity 12 + log(O/H) = 9.04}.
The Milky Way-type disk galaxy is assumed to interact with a companion galaxy with
the same total mass represented by a point-mass particle. The orbit of the two interacting
galaxies is assumed to be hyperbolic with an initial distance
of 280 kpc, orbital eccentricity of 1.1
and pericenter distance of 70 kpc.
We select  TDG candidates from the remnants of the interacting
Milky Way-type disks as follows. We identify the newly formed stars in our simulation.  For each new star we determine the number of additional new stars within 1 kpc of it, N$_{ns}$. For a region with N$_{ns} >$ 100, the total mass of the region is M 
{$\geq$} 3 x 10$^7$ M$_\odot$, so the stars in that region are considered to belong to a TDG candidate.
For each selected TDG candidate, the center-of-mass is estimated
by using all new stars within 1 kpc of the new star. The total mass and the mean gas-phase metallicity
within 1kpc from the center-of-mass are then calculated for the TDG candidate.
We assumed an $R$-band mass to light ratio $M/L_R$ = 0.86 for a stellar population of mean age 1 Gyr and solar metallicity, using the MILES code by \citet{Vazdekis2010}. 
Our simulated data are shown in Figure~\ref{LZ} as black, filled squares. The mean simulated metallicity is 8.57 $\pm$ 0.03, within 3$\sigma$ of the mean observed TDG candidate metallicity of 8.70 $\pm$ 0.05.

\subsection{Choir dwarf galaxies: tidal dwarf candidates, normal dwarfs, and very metal poor dwarfs}
 The Choir giants (M$_R \lesssim $ -20) are in reasonable agreement with the SDSS giant sub-population, having the same metallicity, though being around one magnitude more luminous. The medium-luminosity Choir dwarfs (-16 $\lesssim$ M$_R \lesssim $-20) are also mostly consistent with the SDSS medium-dwarf sub-population. The contours provide a simple diagnostic of the significance of any outlying results. For example, we consider that the two most metal-rich dwarfs at M$_R\sim$ -17.5, being more than 3$\sigma$ from the mean SDSS medium-dwarf population, are bona-fide TDG candidates.
 
 Compared with each of the additional samples of dwarfs listed above, Choir galaxies have an increased scatter at the low luminosity end, spanning the full 1.5 dex metallicity range observed for all types of dwarfs. Some groups (e.g. HIPASS J0400-52) even span this range. The size of the error bars compared with the scatter suggests that this is not a measurement error, but either due to the calibration model or a true dispersion in the population. Figure 3 of \citet{Dopita2013} illustrates with \citet{VanZee1998} H{\sc ii} regions an increased scatter in metallicities between 8.0 $\lesssim$ 12 + log(O/H) $\lesssim$ 8.5 measured using this model. The metallicity calibration for these galaxies depends very strongly on the calibration between log(N/O) and log(O/H), as demonstrated in that figure. However, the observed dispersion is much greater in the Choir sample than in the \citet{VanZee1998} and other comparison samples, so we consider that the Choir dwarf galaxy population is inherently dispersed. We expect that this is due to a wide variation in gas content and environment (distance to host) of the Choir member galaxies. 

We consider that (i) the three Choir dwarfs (J0205-55:S7, J0400-52:S8, J0400-52:S9) with metallicity above 12 + log(O/H) = 8.6 (3-$\sigma$ above the SDSS dwarfs) and above the SDSS 3-$\sigma$ medium-dwarf ellipse are strong TDG candidates (these candidates represent 16\% of the Choir dwarfs fainter than M$_R = -18$), (ii) the dwarfs between 8.0 $\lesssim$ 12 + log(O/H) $\lesssim$ 8.6 and within the SDSS 3-$\sigma$ dwarf ellipse are most likely normal galaxies but still are possible TDGs, and (iii) the dwarfs (J0400-52:S2, J1051-17:S4, J1403-06:S3, J1403-06:S4) with metallicity less than 8.0 are probably the most gas-rich in our sample due to their similarity (in metallicity-luminosity space) to the gas-rich Hydra dwarfs in \citet{Duc2001} and the gas-rich HIPASS J1609-04. The very metal poor sample represents 21\% of the Choir dwarfs fainter than M$_R = -18$. It could be that these very metal-poor dwarfs in our sample have acquired large amounts of relatively pristine H{\sc i} gas and have had star formation re-ignited, due to recent interactions with the nearby giant galaxies \citep[e.g.,][]{Kannappan2013}. We note that \citet{Skillman2013} suggest the 
{most metal-poor} dwarfs may become metal poor due to H{\sc i} gas inflowing to their central star-forming regions.
Our TDG candidates and metal-poor dwarfs are noted in Table~\ref{fluxes}.

We are conducting a follow-up analysis of the kinematics and masses of our TDG candidates to determine which are bona-fide TDGs, and a further analysis of the importance of environmental location and gas content (Sweet et al., in prep).

\section{Conclusions and Future Work}
In this paper we have used the new \citet{Dopita2013} metallicity calibrations to calibrate the luminosity-metallicity relation for a range of galaxy types. Importantly, we used the same calibration for our population of galaxies in H{\sc i}-rich groups as for our control samples.

We make the following points:
\begin{enumerate}
\item In metallicity-luminosity space we find two sub-populations, or clusters, within the SDSS sample. The cluster of metal-rich giants represents 52\% of the sample, while the remaining 48\% are metal-poor dwarfs.
\item There is an apparent floor to the metallicity of SDSS dwarfs at 12 + log(O/H) = 7.9; the mean metallicity for SDSS dwarfs fainter than M$_R$ = -16 is 12 + log(O/H) = 8.28 $\pm$ 0.10.
\item Isolated dwarf galaxies  
{appear to} have a constant metallicity with magnitude of 12 + log(O/H) = 8.46 $\pm$ 0.04, similar to the SDSS dwarf sample.
\item On average, TDG candidates from the literature have a metallicity of 12 + log(O/H) = 8.70 $\pm$ 0.05, significantly elevated above SDSS galaxies. Our simulated TDGs are slightly less metal-rich than TDG candidates in the literature at 12 + log(O/H) = 8.57 $\pm$ 0.03, but still significantly more metal-rich than typical dwarfs.
\item Gas-rich cluster dwarfs 
{trend towards lower metallicy than their isolated counterparts, where Hydra dwarfs }from \citet[][]{Duc2001} have the lowest metallicity of our comparison samples, suggesting that dwarf metallicity is highly dependent 
{on group} membership.
\item At medium-bright magnitudes, our sample of star-forming galaxies in groups is consistent with SDSS.
\item At faint luminosity there is an increase in dispersion in metallicity in our sample, indicating a wide range of H{\sc i} content and environmental location.
\item Based on metallicity, we identify three (16\% of dwarfs) strong TDG candidates (12+log(O/H) $>$ 8.6), which have metallicities consistent with other TDG candidates in the literature and our simulations, and significantly above the SDSS control sample at 12 + log(O/H) = 8.28 $\pm$ 0.10. These galaxies are J0205-55:S7, J0400-52:S8, J0400-52:S9, and are discussed very briefly in Appendix A.1.
\item We also identify four (21\%) very metal-poor galaxies (12+log(O/H) $<$ 8.0), consistent with gas-rich cluster dwarfs whose star formation has been ignited due to interactions with nearby giant galaxies. These galaxies are J0400-52:S2, J1051-17:S4, J1403-06:S3, J1403-06:S4, and are discussed very briefly in Appendix A.2.
\\
\end{enumerate}

To conclude, metallicity can be an important diagnostic for identifying preliminary populations of candidate TDGs
{. Other factors such as environment, as noted by} \citet{Vaduvescu2007}
{ may also influence metallicity, so} careful follow-up is required before declaring the candidates to be bona-fide TDGs.

\section*{Acknowledgements}

{The authors would like to thank the anonymous referee for a constructive report, which improved the clarity of this paper.}

We thank David Rohde for helpful discussions about Gaussian mixture modelling, and Jeff Rich, Mort Canty and James R A Davenport for providing IDL routines.
We acknowledge funding support from the UQ-UWA Bilateral Research Collaboration Award.

This research has made use of NASA's Astrophysics Data System.

This research has made use of the NASA/IPAC Extragalactic Database (NED) which is operated by the Jet Propulsion Laboratory, California Institute of Technology, under contract with the National Aeronautics and Space Administration.

\appendix
\section{Notes on strong TDG candidates and very metal-poor dwarfs}

\subsection{Strong tidal dwarf galaxy candidates}

\subsubsection{HIPASS J0205-55:S7} This dwarf is located in Choir group HIPASS J0205-55, which appears to be comprised of two merging systems \citep{Sweet2013}. The nearest bright neighbour to S7 is S2, approximately 50 kpc away in projection. S7 has the morphology of a symmetric, edge-on disk galaxy with a half-light radius of 0.85 kpc.

\subsubsection{HIPASS J0400-52:S8 and S9}
These dwarfs are very compact with half-light radii of 1.5 and 2.5 kpc respectively. They are very close companions to S4 and S6 at only 20 kpc away in projection from their respective nearest giant galaxies. This Choir group is part of Abell 3193.

\subsection{Very metal-poor dwarfs}
\subsubsection{HIPASS J0400-52:S2}
This dwarf, approximately 100 kpc from its nearest neighbour S1, has a half-light radius of 1.3 kpc, and is comprised of two abutting low surface brightness knots.

\subsubsection{HIPASS J1051-17:S4}
S4 is located about 100 kpc from giant galaxy S1 in the direction of the second-brightest spiral in the group, S2. It may have gained some HI gas and had star formation reignited during a recent encounter with S1 (Kilborn et al., in prep.). It consists of two faint H{\sc ii} regions in a low surface brightness host with a half-light radius of 1.4 kpc.

\subsubsection{HIPASS J1403-06:S3 and S4}
These two quite compact dwarfs are the two faintest members in this Choir group. Both are located about 50 kpc from the nearby giant S1, which itself is currently interacting with the other giant in the group, S2. The interacting pair S1 and S2 is known as Arp 271. The two dwarfs each consist of a single H{\sc ii} region in a low surface brightness host, with half-light radii of 0.7 and 1.2 kpc for S3 and S4 respectively.
    
\clearpage
\addcontentsline{toc}{chapter}{Bibliography}
\bibliographystyle{hapj}
\bibliography{ChoirsLZ}

\begin{centering}
\begin{longtable}{|l|l|l|l|l|l|l|l|l|l|l|l|l|l|}
\caption{WiFeS observations\label{observations}}\\
\hline 
\multicolumn{1}{|c|}{HIPASS+}&\multicolumn{1}{c|}{Optical ID}&\multicolumn{1}{c|}{RA}&\multicolumn{1}{c|}{Dec}&\multicolumn{1}{c|}{Obs. date}&\multicolumn{1}{c|}{Int. time}&\multicolumn{1}{c|}{Mode}\\
\multicolumn{1}{|c|}{}&\multicolumn{1}{c|}{}&\multicolumn{1}{c|}{[h m s]}&\multicolumn{1}{c|}{[d m s]}&\multicolumn{1}{c|}{}&\multicolumn{1}{c|}{[s]}&\multicolumn{1}{c|}{}\\
\multicolumn{1}{|c|}{(1)}&\multicolumn{1}{c|}{(2)}&\multicolumn{1}{c|}{(3)}&\multicolumn{1}{c|}{(4)}&\multicolumn{1}{c|}{(5)}&\multicolumn{1}{c|}{(6)}&\multicolumn{1}{c|}{(7)}\\ 
\hline
\endfirsthead
\hline
J0205-55:S1b& ESO153-G017                    & 02 05 05.48 & -55 06 42.54 &2011/09/20&3600&N\\
J0205-55:S2c& ESO153-IG016                   & 02 04 50.78 & -55 13 01.55 &2011/09/21&5400&N\\
J0205-55:S2d&& 02 04 50.78 & -55 13 01.55 &2011/09/21&5400&N\\
J0205-55:S3a& ESO153-G015                    & 02 04 34.92 & -55 07 09.65 &2012/10/06&3600&N\\
J0205-55:S4a& ESO153-G013                    & 02 04 19.75 & -55 13 50.44 &2011/09/20&2700&N\\
J0205-55:S5         & APMUKS& 02 04 54.77 & -55 08 31.99 &2011/09/12&3600&S\\
J0205-55:S6         & APMUKS& 02 04 57.07 & -55 13 34.10 &2011/09/22&4500&S\\
J0205-55:S7         & 6dF& 02 05 00.57 & -55 15 19.63 &2011/09/22&4500&S\\
J0205-55:S8         & APMUKS& 02 04 29.71 & -55 12 56.09 &2012/10/06&3600&S\\
J0205-55:S9& APMUKS& 02 05 23.76 & -55 14 14.20 &2012/10/07&4800&S\\
J0258-74:S1b     & ESO031-G005                    & 02 58 06.48 & -74 27 22.79 &2012/10/08&2700&N\\
J0258-74:S2          & MRSS& 02 58 52.43 & -74 25 53.25 &2012/10/08&2700&N\\
J0258-74:S3          & 2MASX& 02 58 42.76 & -74 26 03.55 &2012/10/09&3150&N\\
J0258-74:S4          & MRSS& 02 57 29.23 & -74 22 34.75 &2012/10/09&4500&S\\
J0400-52:S1          & ESO156-G029                    & 04 00 40.82 & -52 44 02.71 &2012/10/07&3600&N\\
J0400-52:S2          & APMUKS& 04 00 48.07 & -52 41 02.81 &2012/10/07&3600&S\\
J0400-52:S3          & 2MASX  & 04 00 06.03 & -52 39 32.63 &2012/10/07&2400&S\\
J0400-52:S4          & IC2028       & 04 01 18.23 & -52 42 27.08 &2012/07/08&3150&N\\
J0400-52:S5          & 2MASX& 04 00 53.00 & -52 49 38.43 &2012/07/08&3150&N\\
J0400-52:S6          & IC2029      & 04 01 17.84 & -52 48 02.81 &2012/10/09&4500&N\\
J0400-52:S7          & APMUKS& 04 01 08.99 & -52 49 32.78 &2012/10/09&4500&S\\
J0400-52:S8& -      & 04 01 17.00 & -52 42 08.50 &2012/07/08&3150&N\\
J0400-52:S9& -      & 04 01 19.29 & -52 47 56.10 &2012/10/09&4500&N\\
J1051-17:S1a& 2MASX& 10 51 37.45 & -17 07 29.23 &2011/04/30&1800&C\\
J1051-17:S1b&& 10 51 37.45 & -17 07 29.23 &2011/04/30&2100&C\\
J1051-17:S1c&& 10 51 37.45 & -17 07 29.23 &2011/04/30&2700&C\\
J1051-17:S2a        & NGC3431                        & 10 51 15.11 & -17 00 29.44 &2011/05/01&1800&C\\
J1051-17:S2b&& 10 51 15.11 & -17 00 29.44 &2011/05/01&1800&C\\
J1051-17:S3          & -                       & 10 51 35.94 & -16 59 16.80 &2011/04/28-29&7200&N\\
J1051-17:S4          & -                       & 10 51 26.01 & -17 05 03.61 &2011/04/29&5400&S\\
J1051-17:S5          & -                       & 10 51 50.91 & -16 58 31.64 &2011/05/01&5400&S\\
J1051-17:S6          & -                       & 10 51 42.78 & -17 06 34.59 &2011/04/30&3600&C\\
J1051-17:S7          & -                       & 10 51 33.36 & -17 08 36.63 &2011/04/30&3600&C\\
J1051-17:S8   & -      & 10 51 25.92 & -17 08 16.44 &2011/04/29&5400&S\\
J1051-17:S9   & -      & 10 51 56.54 & -17 05 03.50 &2012/05/18&3600&N\\
J1403-06:S1a& NGC5426                        & 14 03 24.88 & -06 04 09.14 &2012/05/21&1650&N\\
J1403-06:S2a          & NGC5427                        & 14 03 26.09 & -06 01 51.20 &2012/05/21&1800&N\\
J1403-06:S2b&& 14 03 26.09 & -06 01 51.20 &2012/05/21&1800&N\\
J1403-06:S3          & APMUKS& 14 03 13.48 & -06 06 24.17 &2012/05/21&3600&S\\
J1403-06:S4          & APMUKS& 14 03 34.62 & -06 07 59.27 &2012/05/21&5400&S\\
J1408-21:S1a& ESO578-G026                    & 14 08 42.04 & -21 35 49.82 &2012/05/20&2700&N\\
J1408-21:S1b&& 14 08 42.04 & -21 35 49.82 &2012/05/20&5400&N\\
J1408-21:S1c&& 14 08 42.04 & -21 35 49.82 &2012/05/20&4500&N\\
J1408-21:S1d&& 14 08 42.04 & -21 35 49.82 &2012/05/20&4500&N\\
J1408-21:S2          & 2MASX& 14 08 57.72 & -21 38 52.47 &2012/05/18&2700&N\\
J1408-21:S3          & 2MASX& 14 08 41.04 & -21 37 40.97 &2012/05/19&3600&N\\
J1408-21:S4          & 2MASX    & 14 08 33.28 & -21 36 07.18 &2012/05/19&3600&N\\
J1408-21:S5& -      & 14 08 39.82 & -21 38 14.30 &2012/05/19&9000&S\\
J1408-21:S6& -      & 14 08 52.84 & -21 42 07.20 &2012/05/18&3000&S/N\\
J1956-50:S1b&& 19 56 45.51 & -50 03 20.29 &2011/09/20&3600&N\\
J1956-50:S1c&& 19 56 45.51 & -50 03 20.29 &2011/09/20&5400&N\\
J1956-50:S1d& IC4909                         & 19 56 45.51 & -50 03 20.29 &2011/09/22&5400&N\\
J1956-50:S1e&& 19 56 45.51 & -50 03 20.29 &2011/09/22&5400&N\\
J1956-50:S2          & 2MASX     & 19 55 53.21 & -50 02 10.82 &2011/09/21&3600&N\\
J1956-50:S3          & -                       & 19 56 08.20 & -50 02 21.56 &2011/09/21&5400&S\\
J1956-50:S4& -      & 19 55 45.92 & -50 06 15.50 &2011/09/22&5400&S\\
J2027-51:S1a& ESO234-G032                    & 20 28 06.39 & -51 41 29.83 &2011/04/30&3600&C\\
J2027-51:S1b&& 20 28 06.39 & -51 41 29.83 &2011/04/29-30&4200&C\\
J2027-51:S1c&& 20 28 06.39 & -51 41 29.83 &2011/04/30&3600&C\\
J2027-51:S2a      & ESO234-G028                    & 20 27 31.97 & -51 39 20.81 &2011/09/19&3600&N\\
J2027-51:S2c&& 20 27 31.97 & -51 39 20.81 &2011/09/19&3600&N\\
J2027-51:S3          & MRSS& 20 27 48.52 & -51 44 19.35 &2011/04/28&6600&N\\
J2027-51:S4          & -                       & 20 27 54.64 & -51 38 04.52 &2011/04/29&5400&S\\
J2318-42a:S1c& NGC7582                        & 23 18 23.44 & -42 22 11.94 &2012/05/21&1350&N\\
J2318-42a:S1d&& 23 18 23.44 & -42 22 11.94 &2012/05/21&900&N\\
J2318-42a:S2a         & NGC7590                        & 23 18 54.78 & -42 14 18.94 &2012/05/20&1800&N\\
J2318-42a:S2b&& 23 18 54.78 & -42 14 18.94 &2012/05/20&1800&N\\
J2318-42a:S3a& NGC7599                        & 23 19 21.14 & -42 15 24.6 &2012/05/21&1800&N\\
J2318-42a:S3b&& 23 19 21.14 & -42 15 24.6 &2012/05/21&1800&N\\
J2318-42a:S4& APMUKS  & 23 18 50.44 & -42 23 50.30 &2012/05/19&5400&S\\
\hline
\end{longtable}
\end{centering}
\justifying{Columns: (1): SINGG name with (a-e) appended for pointing where applicable; (2): name assigned to group as found in NASA/IPAC Extragalactic Database (NED; http://ned.ipac.caltech.edu/); (3): J2000 right ascension of brightest source in field; (4): J2000 declination of brightest source in field; (5): Date of observations; (6): Total integration time; (7): Mode of observation. N denotes nod and shuffle, S denotes sub-aperture nod and shuffle, C denotes classical observation.}

\tiny
\begin{centering}
\begin{longtable}{|l|l|l|l|l|l|l|l|l|l|l|l|l|l|}
\caption{Measured emission line fluxes and other quantities\label{fluxes}}\\
\hline 
\multicolumn{1}{|c|}{HIPASS+}&\multicolumn{1}{|c|}{H$\beta$}&\multicolumn{1}{|c|}{[O {\sc iii}] }&\multicolumn{1}{|c|}{H$\alpha$}&\multicolumn{1}{|c|}{[N {\sc ii}] }&\multicolumn{1}{|c|}{[S {\sc ii}] }&\multicolumn{1}{|c|}{[S {\sc ii}]}&\multicolumn{1}{|c|}{12+log(O/H)}&\multicolumn{1}{|c|}{M$_R$}&\multicolumn{1}{|c|}{V$_{hel}$}\\
\multicolumn{1}{|c|}{}&\multicolumn{1}{|c|}{4861.3}&\multicolumn{1}{|c|}{5006.9}&\multicolumn{1}{|c|}{6562.8}&\multicolumn{1}{|c|}{6583.4}&\multicolumn{1}{|c|}{6716.4}&\multicolumn{1}{|c|}{6730.8}&\multicolumn{1}{|c|}{}&\multicolumn{1}{|c|}{[mag]}&\multicolumn{1}{|c|}{[km s$^{-1}$]}\\
\multicolumn{1}{|c|}{(1)}&\multicolumn{1}{|c|}{(2)}&\multicolumn{1}{|c|}{(3)}&\multicolumn{1}{|c|}{(4)}&\multicolumn{1}{|c|}{(5)}&\multicolumn{1}{|c|}{(6)}&\multicolumn{1}{|c|}{(7)}&\multicolumn{1}{|c|}{(8)}&\multicolumn{1}{|c|}{(9)}&\multicolumn{1}{|c|}{(10)}\\
\hline
\endfirsthead
\hline
J0205-55:S1 &1560$\pm$140&337$\pm$85&6000$\pm$378&3110$\pm$200&531$\pm$51&371$\pm$23&9.35$\pm^{0.03}_{0.03}$&-22.57$\pm$0.22&6490\\
J0205-55:S2 &9110$\pm$810&13500$\pm$1200&29300$\pm$3000&4590$\pm$490&5120$\pm$490&3630$\pm$340&8.61$\pm^{0.07}_{0.07}$&-19.22$\pm$0.00&5941\\
J0205-55:S3 &423$\pm$119&436$\pm$116&2680$\pm$340&1390$\pm$190&348$\pm$80&243$\pm$30&9.20$\pm^{0.05}_{0.06}$&-22.00$\pm$0.09&6074\\
J0205-55:S4 &417$\pm$64&425$\pm$53&1740$\pm$113&943$\pm$62&665$\pm$46&465$\pm$24&8.94$\pm^{0.04}_{0.04}$&-17.32$\pm$0.18&5941\\
J0205-55:S5***&1100$\pm$390&873$\pm$369&1220$\pm$290&154$\pm$348&0$\pm$0&0$\pm$0&0.00$\pm^{0.00}_{0.00}$&-18.40$\pm$0.06&6216\\
J0205-55:S6          &629$\pm$88&1470$\pm$180&2220$\pm$240&339$\pm$42&563$\pm$55&394$\pm$35&8.41$\pm^{0.09}_{0.09}$&-15.78$\pm$1.08&5758\\
J0205-55:S7*&186$\pm$139&951$\pm$170&855$\pm$141&413$\pm$108&206$\pm$87&144$\pm$16&8.93$\pm^{0.14}_{0.19}$&-17.65$\pm$0.05&5758\\
J0205-55:S8          &1420$\pm$190&738$\pm$190&1750$\pm$270&369$\pm$129&327$\pm$61&229$\pm$32&8.78$\pm^{0.19}_{0.20}$& -15.37$\pm$0.12 &5891\\
J0205-55:S9 &399$\pm$28&1520$\pm$140&1330$\pm$200&93$\pm$12&167$\pm$23&117$\pm$16&8.28$\pm^{0.22}_{0.26}$&-21.55$\pm$0.12&6120\\
J0258-74:S1    &1290$\pm$130&789$\pm$75&5760$\pm$610&1880$\pm$220&815$\pm$84&596$\pm$56&9.08$\pm^{0.05}_{0.06}$&-19.48$\pm$0.29&4883\\
J0258-74:S2           &864$\pm$99&2560$\pm$290&2920$\pm$440&250$\pm$41&371$\pm$58&260$\pm$40&8.40$\pm^{0.12}_{0.13}$&-18.51$\pm$0.56&4883\\
J0258-74:S3           &451$\pm$45&226$\pm$24&2620$\pm$230&983$\pm$89&668$\pm$61&505$\pm$46&8.98$\pm^{0.05}_{0.05}$&-17.32$\pm$0.85&4655\\
J0258-74:S4           &1020$\pm$100&3700$\pm$340&3490$\pm$510&175$\pm$29&403$\pm$59&303$\pm$42&8.11$\pm^{0.14}_{0.14}$&-21.90$\pm$0.06&4838\\
J0400-52:S1***&280$\pm$31&4720$\pm$520&0$\pm$0&0$\pm$0&0$\pm$0&0$\pm$0&0.00$\pm^{0.00}_{0.00}$&-15.91$\pm$0.02&10424\\
J0400-52:S2**&271$\pm$31&815$\pm$85&1150$\pm$120&16$\pm$9&90$\pm$13&63$\pm$6&7.27$\pm^{0.70}_{1.60}$&-20.02$\pm$0.01&11659\\
J0400-52:S3           &1900$\pm$620&315$\pm$177&7570$\pm$830&3020$\pm$290&1140$\pm$150&794$\pm$59&9.23$\pm^{0.07}_{0.07}$&-22.82$\pm$0.01&11384\\
J0400-52:S4           &305$\pm$31&159$\pm$33&1430$\pm$94&698$\pm$53&204$\pm$28&154$\pm$7&9.21$\pm^{0.04}_{0.04}$&-22.71$\pm$0.05&9967\\
J0400-52:S5           &1350$\pm$120&420$\pm$68&6740$\pm$300&3100$\pm$140&905$\pm$60&904$\pm$34&9.22$\pm^{0.02}_{0.02}$&-22.13$\pm$0.06&10790\\
J0400-52:S6           &171$\pm$29&81$\pm$24&1240$\pm$130&476$\pm$49&189$\pm$22&132$\pm$1&9.15$\pm^{0.05}_{0.05}$&-18.89$\pm$0.13&10287\\
J0400-52:S7           &526$\pm$105&456$\pm$114&1740$\pm$180&423$\pm$47&463$\pm$63&337$\pm$27&9.08$\pm^{0.11}_{0.11}$& -17.36$\pm$0.08 &10607\\
J0400-52:S8*&51$\pm$22&52$\pm$32&252$\pm$38&119$\pm$31&72$\pm$13&51$\pm$8&8.70$\pm^{0.08}_{0.08}$& -17.56$\pm$0.07 &9921\\
J0400-52:S9*&89$\pm$19&53$\pm$48&728$\pm$72&365$\pm$49&193$\pm$34&135$\pm$9&8.98$\pm^{0.13}_{0.16}$&-22.16$\pm$0.21&10287\\
J1051-17:S1 &16$\pm$11&31$\pm$11&46$\pm$11&8$\pm$8&4$\pm$7&3$\pm$0&8.96$\pm^{0.02}_{0.03}$&-21.55$\pm$0.27&5465\\
J1051-17:S2       &11$\pm$2&14$\pm$2&139$\pm$11&135$\pm$10&40$\pm$3&30$\pm$2&9.21$\pm^{0.03}_{0.03}$&-21.85$\pm$0.05&5288\\
J1051-17:S3           &167$\pm$72&255$\pm$82&273$\pm$75&101$\pm$49&29$\pm$16&48$\pm$10&9.00$\pm^{0.19}_{0.28}$&-16.34$\pm$0.09&5969\\
J1051-17:S4**&94$\pm$17&196$\pm$29&394$\pm$54&15$\pm$8&67$\pm$11&47$\pm$6&7.50$\pm^{0.41}_{0.74}$&-17.20$\pm$0.06&5465\\
J1051-17:S5           &35$\pm$26&60$\pm$30&1540$\pm$220&176$\pm$33&373$\pm$51&261$\pm$30&8.40$\pm^{0.14}_{0.15}$&-16.95$\pm$0.04&5465\\
J1051-17:S6           &30$\pm$4&55$\pm$7&121$\pm$15&17$\pm$2&19$\pm$2&13$\pm$2&8.60$\pm^{0.09}_{0.09}$&-16.94$\pm$0.12&5648\\
J1051-17:S7           &9$\pm$1&11$\pm$1&32$\pm$4&3$\pm$0&5$\pm$1&4$\pm$0&8.42$\pm^{0.10}_{0.11}$& -17.48$\pm$0.04 &5374\\
J1051-17:S8    &28$\pm$9&73$\pm$17&374$\pm$44&200$\pm$24&166$\pm$20&134$\pm$15&8.80$\pm^{0.08}_{0.08}$& -16.68$\pm$0.05 &5294\\
J1051-17:S9    &99$\pm$29&109$\pm$26&507$\pm$70&131$\pm$27&176$\pm$27&123$\pm$15&8.63$\pm^{0.13}_{0.13}$&-23.05$\pm$0.20&5582\\
J1403-06:S1 &1290$\pm$90&301$\pm$37&3110$\pm$210&1350$\pm$90&489$\pm$39&359$\pm$24&9.20$\pm^{0.03}_{0.03}$&-22.74$\pm$0.09&2498\\
J1403-06:S2         &1780$\pm$160&12900$\pm$900&4820$\pm$460&5470$\pm$500&1310$\pm$120&1340$\pm$120&9.10$\pm^{0.03}_{0.04}$&-22.74$\pm$0.10&2727\\
J1403-06:S3**&600$\pm$51&1460$\pm$130&1200$\pm$170&64$\pm$12&209$\pm$31&146$\pm$21&7.93$\pm^{0.17}_{0.20}$&-14.51$\pm$0.86&2753\\
J1403-06:S4**&208$\pm$28&233$\pm$27&520$\pm$86&13$\pm$7&78$\pm$21&55$\pm$9&7.40$\pm^{0.58}_{1.39}$&-23.15$\pm$0.26&2671\\
J1408-21:S1 &2160$\pm$290&878$\pm$130&9140$\pm$540&4600$\pm$280&1080$\pm$90&966$\pm$75&9.25$\pm^{0.03}_{0.03}$&-23.15$\pm$0.27&8694\\
J1408-21:S2           &661$\pm$76&212$\pm$46&2780$\pm$189&1050$\pm$70&404$\pm$37&352$\pm$19&9.15$\pm^{0.04}_{0.04}$&-20.72$\pm$0.10&8821\\
J1408-21:S3           &887$\pm$86&258$\pm$34&3770$\pm$310&1330$\pm$120&637$\pm$62&505$\pm$30&9.10$\pm^{0.04}_{0.05}$&-21.23$\pm$0.08&8782\\
J1408-21:S4           &202$\pm$25&104$\pm$31&998$\pm$104&393$\pm$41&237$\pm$27&165$\pm$14&9.02$\pm^{0.06}_{0.07}$& -13.31$\pm$0.52 &9126\\
J1408-21:S5 &204$\pm$402&428$\pm$601&654$\pm$284&101$\pm$228&119$\pm$94&83$\pm$13&8.57$\pm^{0.13}_{0.14}$& -17.12$\pm$0.07  &8778\\
J1408-21:S6***&336$\pm$155&15$\pm$124&329$\pm$127&0$\pm$0&0$\pm$0&0$\pm$0&0.00$\pm^{0.00}_{0.00}$&-22.35$\pm$0.19&8672\\
J1956-50:S1 &262$\pm$61&218$\pm$56&1570$\pm$180&631$\pm$83&172$\pm$34&126$\pm$15&9.18$\pm^{0.05}_{0.06}$&-22.35$\pm$0.20&7610\\
J1956-50:S2           &2170$\pm$2300&1710$\pm$190&7670$\pm$720&2110$\pm$200&1510$\pm$150&1080$\pm$100&8.89$\pm^{0.06}_{0.06}$&-16.65$\pm$0.22&7015\\
J1956-50:S3           &1590$\pm$160&10700$\pm$900&5070$\pm$610&110$\pm$20&198$\pm$31&139$\pm$20&8.24$\pm^{0.16}_{0.15}$& -15.15$\pm$0.23 &6375\\
J1956-50:S4 &242$\pm$31&413$\pm$55&610$\pm$82&59$\pm$14&94$\pm$14&70$\pm$9&8.40$\pm^{0.16}_{0.18}$&-21.73$\pm$0.29&7472\\
J2027-51:S1 &51$\pm$5&47$\pm$4&621$\pm$33&321$\pm$16&134$\pm$8&226$\pm$14&8.99$\pm^{0.03}_{0.03}$&-21.73$\pm$0.30&5830\\
J2027-51:S2 &42$\pm$3&29$\pm$2&980$\pm$64&306$\pm$20&155$\pm$10&166$\pm$11&9.01$\pm^{0.04}_{0.04}$&-21.91$\pm$0.17&5783\\
J2027-51:S3           &1300$\pm$140&2740$\pm$280&4240$\pm$550&623$\pm$75&997$\pm$121&708$\pm$86&8.43$\pm^{0.09}_{0.10}$&-17.40$\pm$0.15&5830\\
J2027-51:S4           &1740$\pm$140&3080$\pm$260&5770$\pm$660&815$\pm$99&1130$\pm$130&792$\pm$86&8.50$\pm^{0.08}_{0.09}$&-22.25$\pm$0.26&6013\\
J2318-42a:S1 &8610$\pm$500&25400$\pm$1000&61900$\pm$2100&45100$\pm$1400&10300$\pm$300&9670$\pm$290&9.16$\pm^{0.01}_{0.01}$&-22.25$\pm$0.27&1461\\
J2318-42a:S2  &583$\pm$81&2300$\pm$230&5780$\pm$480&5170$\pm$390&2450$\pm$190&2110$\pm$130&8.99$\pm^{0.04}_{0.04}$&-21.22$\pm$0.06&1481\\
J2318-42a:S3 &2890$\pm$220&1490$\pm$130&13700$\pm$1500&3280$\pm$430&1800$\pm$220&1270$\pm$150&9.01$\pm^{0.07}_{0.07}$&-20.15$\pm$0.29&1777\\
J2318-42a:S4***&88$\pm$24&29$\pm$21&213$\pm$30&222$\pm$15&16$\pm$7&34$\pm$4&0.00$\pm^{0.00}_{0.00}$&&1685\\
\hline
\end{longtable}
\end{centering}
\justifying{Columns: (1): SINGG name; (2-7): observed (and extinction-corrected fluxes only in electronic version) for various emission lines, in units of 10$^{-17}$ erg s$^{-1}$ cm$^{-2}$; (8): metallicity calibrated using \citet{Dopita2013}; (9) SINGG R-band absolute magnitude; (10): WiFeS heliocentric velocity. (*): TDG candidates; (**): very metal-poor dwarfs; (***): metallicity not measurable due to poor signal in one or more lines.}
    
\end{document}